\documentstyle[12pt]{article}
\begin{document}
\begin{center}
{\bf QUANTIZATION OF GRAVITY IN THE THEORY WITH THE MASSIVE
VECTOR FIELD }
\end{center}
\medskip

D.L.Khokhlov
\medskip

Sumy State University, R.-Korsakov St. 2, Sumy 244007 Ukraine

e-mail: others@monolog.sumy.ua

\medskip
\medskip
\medskip
\medskip

\begin{tabular}{p{5.8in}}
{\small
The quantum theory of gravity is considered based on the assumption
that gravitational interaction occurs by means of the vector field
of the Planck mass. Gravitational emission is considered as a
process of the decay of proton into some matter fields at the
Planck scale. Within the framework of grand unification vector
field of the Planck mass may be thought of as those which realize
the interaction between leptons and quarks.
}
\end{tabular}

\medskip
\medskip
\medskip
\medskip
\medskip

In the Einstein theory of gravity~\cite{Buch} with the Lagrangian
\begin{equation}
L=-{1\over{16\pi G}}\sqrt{g}R,\label{eq:a}
\end{equation}
gravitational field is defined by the components of the metric
tensor $g_{\mu\nu}$. If to take $g_{\mu\nu}$ in the form
\begin{equation}
g_{\mu\nu}=\eta_{\mu\nu}+h_{\mu\nu},\label{eq:b}
\end{equation}
where $\eta_{\mu\nu}$ is the background metric and
$h_{\mu\nu}$ is the small perturbations of the background metric,
one comes to the theory of the fields $h_{\mu\nu}$ in
the background metric.
Quantization of gravity reduces to quantization
of the field $h_{\mu\nu}$.
As follows from the theory~\cite{Buch},
the field $h_{\mu\nu}$ is a massless one of the spin 2.

Quantum theory of gravity based on the above approach is
nonrenormalizable. According to the dimensions analysis,
the nonrenormalizable theories have the negative mass dimensions.
The Newton constant $G$ has the dimensions ${\rm [m]^{-2}}$
($\hbar=c=1$).
In the theory of the weak interaction, the Fermi constant $G_F$
also has the dimensions ${\rm [m]^{-2}}$. Renormalization of
the weak interaction is effected within the framework of
the theory with the massive vector boson~\cite{Bog}.
From this, it is natural to consider gravity within the
framework of the theory with the massive vector boson.

Let us assume that gravitational interaction occurs by means of
the vector field, denote it $Pl$.
Let us define the coupling of the theory as
\begin{equation}
g=(\hbar c)^{1/2}.\label{eq:c}
\end{equation}
Then the Newton constant is given by
\begin{equation}
G={g^2\over m_{Pl}^2},\label{eq:d}
\end{equation}
where $m_{Pl}=(\hbar c/G)^{1/2}$ is the Planck mass.
This makes natural to assume that the field $Pl$ has the mass
equal to the Planck mass. One can develop the theory of gravity
with the use of the theory with the massive vector field~\cite{Bog}.
To obtain the renormalizable theory,
the original field $Pl$ is taken to be massless and
acquires mass due to the Higgs mechanism.

According to the standard theory~\cite{Bog},
take the Lagrangian of interaction of the matter fields $\psi$
with the massive vector field $Pl$
in the form
\begin{equation}
L=-gJ_{\mu}(x)Pl_{\mu}(x).\label{eq:m}
\end{equation}
The current $J_{\mu}$ is given by
\begin{equation}
J_{\mu}(x)=\bar\psi(x)O_{\mu}\psi(x),\label{eq:m1}
\end{equation}
where $O_{\mu}$ is some matrices.
Matrix element of the Lagrangian (\ref{eq:m}) is of the order
\begin{equation}
M\sim{g^2\over{m_{Pl}^2-q^2}}.\label{eq:n0}
\end{equation}
In the low energy limit $q^2<<m_{Pl}^2$, this takes the form
\begin{equation}
M\sim{g^2\over m_{Pl}^2}.\label{eq:n}
\end{equation}
Matrix element (\ref{eq:n}) describes the processes of the type
\begin{equation}
\psi_1\psi_2\rightarrow Pl\rightarrow\psi_1\psi_2,\label{eq:l}
\end{equation}
which correspond to the Newton interaction of the fields
$\psi_1$, $\psi_2$.
Also matrix element (\ref{eq:n}) describes the processes of the type
\begin{equation}
\psi_1\psi_2\rightarrow Pl\rightarrow\psi_3\psi_4,\label{eq:l1}
\end{equation}
which correspond to the transformation of the fields
$\psi_1$, $\psi_2$ into the fields $\psi_3$, $\psi_4$.
It is natural to thought of gravitational emission as a process
of the type (\ref{eq:l1}). Then gravitational radiation may be
identified with some matter fields $\psi$.
In view of the theory with the Lagrangian~(\ref{eq:m}),
the Newton potential $\varphi_N$ and consequently the metric
fields $h_{\mu\nu}$ are effective.
From this it follows that the metric fields $h_{\mu\nu}$
are not quantized, and the Einstein theory describes only
the classical gravity.

The theory of gravity with the massive vector field is
consistent with the quantum field theory. This allows to incorporate
gravity in the standard scheme of grand unification.
Let us consider unification of electromagnetic $U(1)$,
weak $SU(2)$
and strong $SU(3)$ interactions within the framework
of the $SU(3)\times SU(2)\times U(1)$ model~\cite{Okun}.
Usually it is supposed that unification of the interactions
leads to the confluence of the electromagnetic $\alpha_e$,
weak $\alpha_w$ and strong $\alpha_s$ couplings into one
coupling of grand unification $\alpha_{GU}$.
That is the strong symmetry is described by one simple group,
e.g. $SU(5)$.
The energy of grand unification $E_{GU}\sim 10^{15}\ {\rm GeV}$
is determined as that when the couplings of all the interactions
are the same. Let us assume that unification of the
interactions is not accompanied with the confluence of the
couplings into one coupling.
That is the strong symmetry is described by
the\break $SU(3)\times SU(2)\times U(1)$ group.
Suppose that the energy of grand unification is the Planck mass
$E_{GU}\sim m_{Pl}\sim 10^{19}\ {\rm GeV}$.
In this case we can identified the field $Pl$ with the fields
$X$,~$Y$ which realize interaction between quarks and leptons.
Then the Lagrangian~(\ref{eq:m}) describes interaction of
quarks and leptons with the fields $X$,~$Y$, with the current~$J_{\mu}$
corresponds to the elementary processes of the type
\begin{equation}
ud\rightarrow X\rightarrow e^{+}\bar u,\quad
ud\rightarrow Y\rightarrow\bar d\bar\nu_{e},\quad
ud\rightarrow Y\rightarrow e^{+}\bar u.
\label{eq:p}
\end{equation}
These lead to the reactions of the decay of proton
\begin{equation}
p\rightarrow e^{+}\pi^0,\ \pi^{+}\bar\nu_{e},\ e^{+}\pi^-\pi^+.
\label{eq:r}
\end{equation}
From this gravitational emission may be identified with
the process of the decay of proton into leptons
at the Planck scale.
Then gravitational radiation may be identified with neutrinos~$\nu$.

\end{document}